\documentstyle[epsf]{mn}
\begin{document}
\input{epsf}
\title[The dust shell around HD 179821]
{Scattered light models of the dust shell around HD 179821}

\author[T.M.~Gledhill \& M.~Takami]
        {T.M.~Gledhill$^{1}$\thanks{email: T.Gledhill@star.herts.ac.uk}
         and M.~Takami$^{1}$ \\
         $^{1}$Department of Physical Sciences, University of
        Hertfordshire, 
        College Lane, Hatfield, \\ Hertfordshire AL10 9AB, England
         }
\maketitle

\begin{abstract}
The dust shell around the evolved star HD 179821 has been detected in
scattered light in near-IR imaging polarimetry observations. Here, we
subtract the contribution of the unpolarized stellar light to obtain
an intrinsic linear polarization of between 30 and 40 per cent in the
shell which seems to increase with radial offset from the star. The $J$
and $K$ band data are modelled using a scattering code to determine the
shell parameters and dust properties. We find that the observations
are well described by a spherically symmetric distribution of dust
with an $r^{-2}$ density law, indicating that when mass-loss was
occurring, the mass-loss rate was constant. The models predict that
the detached nature of a spherically symmetric, optically thin dust
shell, with a distinct inner boundary, will only be apparent in
polarized flux. This is in accordance with the observations of this
and other optically thin circumstellar shells, such as IRAS
17436+5003. By fitting the shell brightness we derive an  optical
depth to the star which is consistent with V band observations and
which, assuming a distance of 6 kpc, gives an inner shell radius of
$r_{\rm in}=1.44\times10^{15}$ m, a dust number density of $N_{\rm
in}=2.70\times10^{-1}$ m$^{-3}$ at $r_{\rm in}$ and a dust mass of
$M_{d}=0.08$~M$_{\odot}$. We have explored axisymmetric shell models but
conclude that any deviations from spherical symmetry in the shell must
be slight, with an equator-to-pole density contrast of less than
2:1. We have not been able to simultaneously fit the high linear
polarizations and the small ($E(J-K)=-0.3$) colour excess of the
shell and we attribute this to the unusual scattering properties of
the dust. We suggest that the dust grains around HD 179821 are either
highly elongated or consist of aggregates of smaller particles.

\end{abstract}


\section{Introduction}

The evolved star HD 179821 (IRAS 19114+0002) is surrounded by a dusty
circumstellar envelope which is detectable in scattered light out to a
radius of at least 9 arcsec (Kastner \& Weintraub 1995; Ueta et
al. 2000). Mid-IR imaging (Hawkins et al. 1995) shows that the warm
dust is distributed in a shell around the star with an inner boundary
at a radius of 1.75 arcsec (Jura \& Werner 1999). There appears to be
little or no dust inside this inner boundary, suggesting that the mass-loss 
terminated some time ago. The detached nature of the shell was
also indicated by earlier ground-based and {\em IRAS} photometry which
showed the double-peaked profile characteristic of emission from a hot
photosphere and warm dust shell (Hrivnak, Kwok \& Volk 1989; hereafter
HKV89).

It is still not clear whether HD 179821 is a low or intermediate mass
post-AGB star or a more massive supergiant (see Reddy \& Hrivnak 1999
for a review of the arguments, also Th\'{e}venin et al. 2000). This
fundamental uncertainty is reflected in the wide range of estimates of
the distance to the star (between 1.5 and 6 kpc) and the quantities
dependent upon it, such as the physical size of the shell and the mass
of dust contained. If the star is in the post-AGB phase then the
object is a proto-planetary nebula (PPN) and similar to other
optically thin shell-like PPNe such as IRAS 17436+5003 (Ueta et
al. 2000; Gledhill et al. 2001)

Near-IR imaging polarimetry of HD 179821 (Gledhill et al. 2001)
allowed the overlying stellar point spread function (PSF) to be
separated from the scattered light to provide an image of the dust
shell in polarized flux in the inner 4 arcsec around the star. These
images show a spatially resolved ring of scattered light encircling
the star, with an inner radius of between 1.1 and 1.6 arcsec (the star
is not at the centre of the shell), indicating that both the near-IR
and mid-IR emission arise from the same body of dust illuminated and
heated by the star. By subtracting the stellar PSF, an image of the
shell surface brightness in the near-IR, and hence the intrinsic shell
polarization, can be formed. In this paper we use these quantities,
along with the distribution of polarized flux, to fit scattering
models to the observations in order to deduce the nebula geometry,
dust distribution and the properties of the dust grains.

\section{Observational Constraints}

\subsection{Imaging polarimetry observations}
\begin{figure}
\label{jpi}
\epsfxsize=12cm \epsfbox[100 157 720 700]{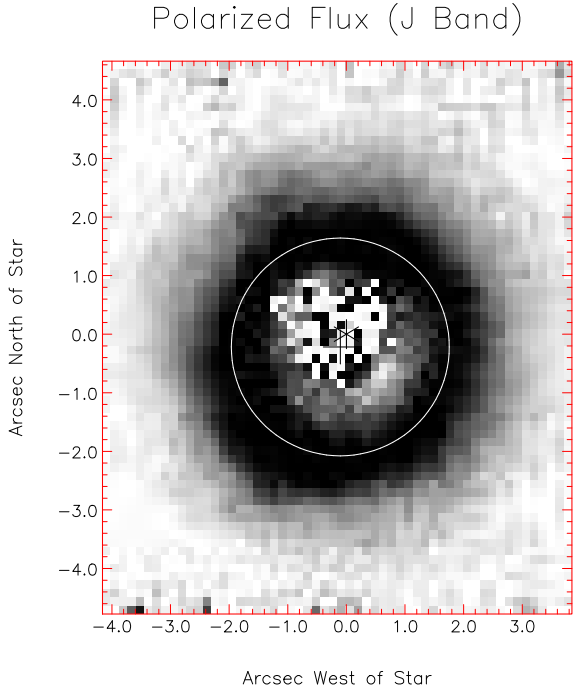}
\caption{A $J$ band image in polarized flux of HD 179821 (IRAS
19114+0002).  The data are from Gledhill et al. (2001). A circle of
radius $1.9$ arcsec, passing through the regions of peak flux,
illustrates the symmetry of the shell. The centre of the circle is
denoted by $+$, which is offset from the stellar centroid, denoted
by $*$.}
\end{figure}
A $J$ band image of HD 179821 in polarized flux, obtained at the 3.8-m UK
Infrared Telescope (UKIRT), is shown in Fig. 1. The polarimetric
observations were made with the IRCAM infrared detector and the IRPOL
polarization module and are described by Gledhill et
al. (2001). A 2x warm magnifier gave a pixel scale of 0.143 arcsec.
Fig. 1 clearly shows the scattered light to lie in a shell
or ring surrounding the star.  A circle of radius $1.9$ arcsec is
drawn on the image to indicate the peak intensity (in polarized flux)
in the shell. Although there is evidence for structure in the
scattered light, to a good approximation the shell is symmetric, with
equal amounts of flux lying inside and outside the circle. The centre
of symmetry ($+$), is offset from the stellar centroid ($*$) which
lies $0.3$ arcsec to the N, indicating that the star does not lie at
the centre of the shell. Jura \& Werner (1999) noted this offset in
their $11.7$~$\mu$m images which show that the thermal emission is also
distributed in a shell around the star, as first observed by Hawkins
et al. (1995). The bright shell seen in polarized flux is surrounded
by a fainter halo of scattered light, which can be seen in our images
out to a radius of approximately $5$ arcsec from the star. In
addition, Fig. 1 shows that polarized flux is recorded inside the
bright shell. The brightness of HD 179821 in the near-IR (K = $4.59
\pm 0.05$, Kastner \& Weintraub 1995) means that alignment and
subtraction residuals dominate the central region of our images, so
that we have not been able to measure polarization closer than 1
arcsec to the star.

\begin{figure}
\label{rawprof}
\epsfxsize=12cm \epsfbox[90 49 643 780]{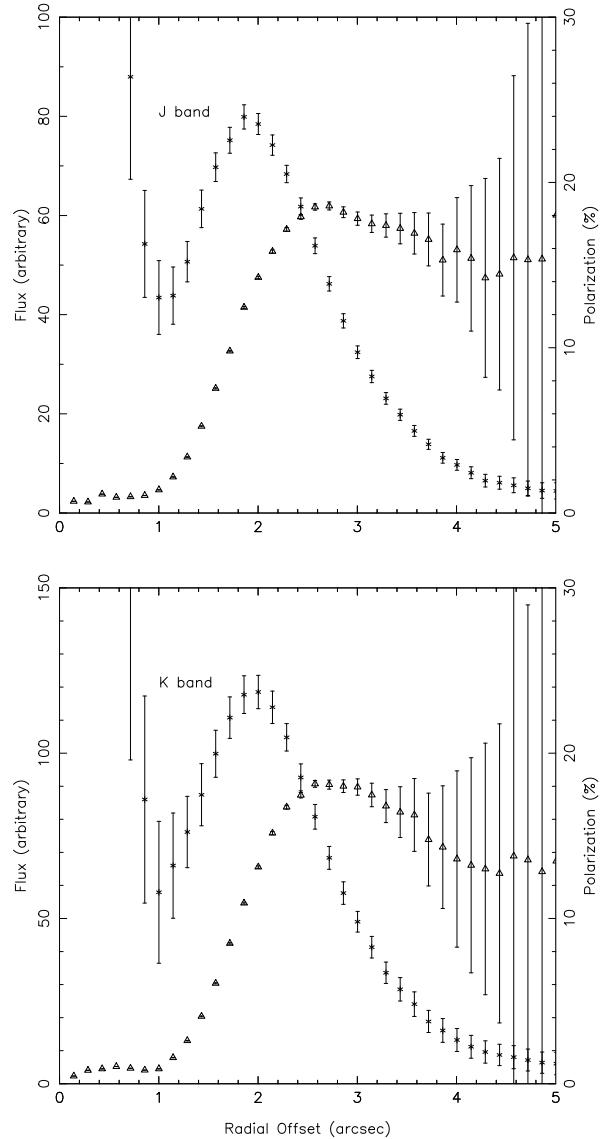}
\caption{Azimuthally averaged radial profiles through the shell in the
$J$ (upper) and $K$ (lower) wavebands. Polarized flux measurements are shown
as crosses and are plotted on an arbitrary flux scale. Per cent polarization
is plotted as triangles according to the scale on the right.}
\end{figure}
We take advantage of the radial symmetry to form azimuthally averaged
radial profiles of polarized flux and per cent polarization, shown in
Fig. 2. The profiles are centred on the $+$ symbol in Fig. 1. Since
the observations were made through varying degrees of cloud cover, it
has not been possible to calibrate the flux scale. The peak in
polarized flux occurs at a radius of $1.9$ arcsec at $J$ and $2.0$
arcsec at $K$. Given that the pixel scale is $0.14$ arcsec and that
the seeing conditions at the time were variable, the difference is not
significant. Per cent polarization peaks at $18.5$ per cent at $J$ and
$18.0$ per cent at $K$, both at a radius of $2.7$ arcsec.  In fact,
the radial profiles in both polarized flux and per cent polarization
are the same, within errors, at $J$ and $K$. This suggests that the
scattering properties of the dust are remarkably wavelength
independent.

\subsection{PSF subtraction}
The polarizations plotted in Fig. 2 were obtained by dividing the
polarized flux (Fig. 1) by the total flux image. However, this
includes a contribution from the unpolarized point spread function
(PSF) of the star, whose wings extend to cover the whole of our
image. These polarizations are, therefore, {\em lower limits} on the
intrinsic polarization of the shell, $P_{\rm int}$. To estimate the 
intrinsic polarization, we
must subtract the PSF of the star to obtain an image of the shell in
total flux so that
\begin{equation}
P_{\rm int} = \frac{\sqrt{Q^{2}+U^{2}}}{I-I_{\rm psf}}
\end{equation}
where, $I$, $Q$, $U$ are the Stokes intensities in the unsubtracted data
and $I_{\rm psf}$ is the subtracted PSF flux. Note that the polarization
angles (and hence the centro-symmetric scattering pattern) depend only on
$Q$ and $U$ and are not affected by the PSF subtraction.

To perform the PSF subtraction we used scaled and aligned images of
nearby bright stars: HD 178744 (B5V, V=6.3) for the $J$ band and HD
179626 (F7V, V=9.1) for $K$ band. Both stars were observed in the same
polarimetric mode as the target for a total of 60 sec each, resulting
in a signal-to-noise ratio (SNR) in the wings of the PSF of greater
than 2 out to a radius of 4 arcsec. The scaling was based on the peak
pixel values of the stars. The PSF subtraction can only be
approximate; seeing conditions at the time were variable and no better
than $\sim 1$ arcsec FWHM. Indeed, in all of our subtractions we found
that the central pixel was surrounded by a negative `hole' indicating
over-subtraction. This suggests that either (i) the alignment between
the target and PSF star could not be achieved to sufficient accuracy
to correct the central bright regions or (ii) the shape of the PSF
star was slightly different in the central peak (broader) compared
with the target. Both could be due to slight variations in seeing. A
reduction in the subtracted flux of $10$ per cent was sufficient to
turn the negative hole into a positive pedestal, so that the central
flux is then under-subtracted. Subtraction of the PSF approximately
doubles the polarization in the shell in both wavebands to between 30
and 40 per cent, so an error in the PSF subtraction of $\pm10$ per
cent represents an error of $\pm5$ per cent in the intrinsic
polarization. The fact that the polarization is observed to double
indicates that although the SNR in the wings of the standard star is
low compared to the target, there is sufficient signal to correct for
the stellar PSF.

After subtraction of the PSF, radial profiles were formed of the total
flux in the shell and of the intrinsic per cent polarization and these are
shown in Fig. 3.
\begin{figure}
\label{subprof}
\epsfxsize=12cm \epsfbox[110 49 663 780]{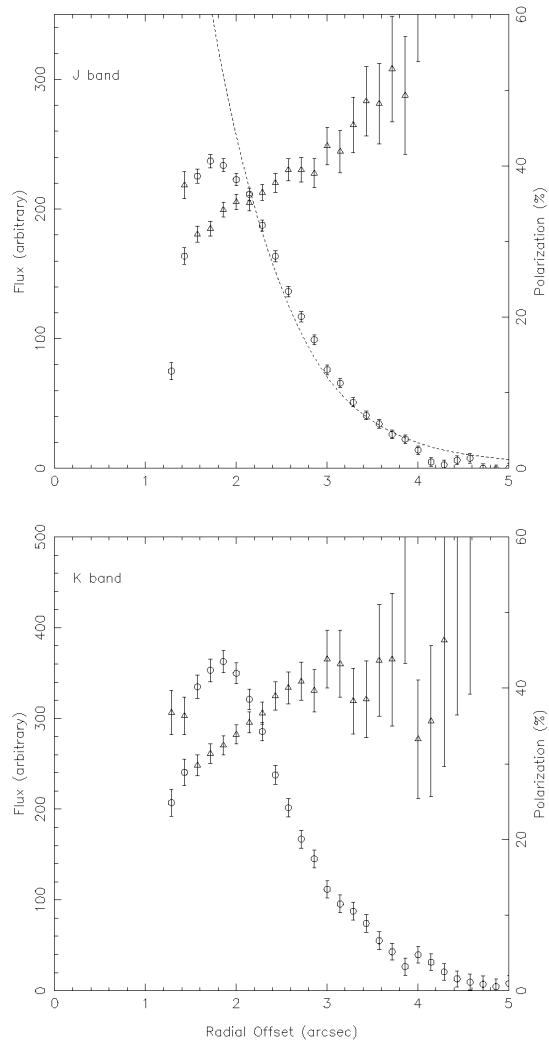}
\caption{Azimuthally averaged radial profiles through the shell in the
$J$ (upper) and $K$ (lower) wavebands, after subtraction of the stellar
PSF.  Total flux measurements are shown as circles and are plotted on
an arbitrary scale. Per cent polarization is plotted as triangles
according to the scale on the right. At radii less than 1 arcsec, the
total flux becomes negative indicating that the PSF has been
over-subtracted in the central regions and consequently the turn over
in total flux at 1.8 arcsec radius may be an artifact. The dashed
curve on the upper plot is V band {\em HST} data (see text).}
\end{figure}
The polarization
now rises with increasing offset from the star, suggesting that the
depolarizing effect of the stellar PSF increases with radius. As with
the polarized flux, the profiles of total flux and per cent
polarization show little variation between $J$ and $K$. The error bars
shown in Fig. 3 are derived from the spread in pixel values during the
azimuthal averaging process and do not include any systematic errors
that may arise from the PSF subtraction (above). In Figure 3 we also plot 
the profile from the V band
{\em HST} image of Ueta et al. (2000)\footnote{courtesy of the NCSA
Astronomy Digital Image Library (ADIL)} which we have obtained by
smoothing their image to the $0.9$ arcsec FWHM seeing of our UKIRT
data, forming the azimuthally averaged profile and then scaling it to
our intensity profile at a radial offset of 1.9 arcsec. Although the
{\em HST} image has not been corrected for the WFPC2 PSF, its
contribution to the shell brightness at several arcsec offset from
the star will be much less than in our ground-based data. The
similarity of the {\em HST} and PSF-subtracted UKIRT profiles gives us
confidence that our PSF subtraction is fairly sound. Indeed, both
profiles approximate an $r^{-3}$ surface brightness dependence, as
expected for optically thin scattering in a shell with an $r^{-2}$
density fall-off. At a radial offset of less that 1.5 arcsec, the
PSF-corrected flux shown in Fig. 3 starts to decrease, and at radii
of less than 1.0 arcsec, it becomes negative. This indicates an
over-subtraction in the central regions.  There is no indication from
the {\em HST} data of a decrease in shell brightness in the central
regions, and so this may be an artifact of the PSF
subtraction.  It is also possible that too much flux has been
subtracted at radii $>1.5$ arcsec and that the intrinsic polarization
is lower than we have estimated. To try to further corroborate the PSF
subtraction, we also used a separate $J$ band observation of HD 179821
and PSF standard, taken in June 1999 (a year after the polarimetry
observations) to form a PSF-subtracted image of the shell. We then
divided this image into our (May 1998) polarized flux image and find
that the shell polarization is the same, within errors, to that shown
in Fig. 3.

\subsection{Shell properties}

\subsubsection{Relative brightness and colour}
A measure of the scattering cross-section, $C_{\rm sca}$, of the dust
in the shell is provided by the ratio of the shell surface brightness
to the brightness of the star. This ratio, which we call $S$, is
intrinsic to the shell and independent of interstellar extinction. The
variation of $S$ with wavelength provides a strong diagnostic of the
dust model, in particular the size distribution. To quantify the
surface brightness of the shell, we use the PSF-subtracted azimuthally
averaged radial profile (Fig. 3) and take the counts in a bin at
radial offset of $2.2$ arcsec. The ratio of this value to the counts
in an equivalent bin at the position of the star in the {\em unsubtracted} 
data gives an
estimate of $S$, the relative brightness of the shell. Note that the
brightness of the shell at the position of the star is negligible
compared to that of the star itself.  We also estimate $S$ for the
$0.55$~$\mu$m WFPC2 observations of Ueta et al. (2000) shown in
Fig. 3. Since this data is not PSF subtracted, $S_{0.55}$ is likely to
be an upper limit on the shell brightness in the V band. We summarize
these quantities, along with the wavelength variation, in Table 1. As
can be seen, there is remarkably little variation in the relative
shell brightness, and hence the scattering efficiency, between $J$ and
$K$. The ratio $S_{J}/S_{K}=1.32$ is much less than the $\lambda^{-4}$
dependence expected for scattering from small (Rayleigh) particles and
gives a colour excess for the shell of $E(J-K)=-0.3$. This `grey'
scattering law was also noted by Kastner \& Weintraub (1995) who
suggested that the grains must be large and/or have unusual scattering
properties.

\begin{table}
\begin{tabular}{llll}
                & $0.55$~$\mu$m  &      $J$         &       $K$           \\ \\
$P_{\lambda}$             & -            & $34.2\pm0.9$   &  $32.5\pm1.3$        \\
$S_{\lambda}\times10^{2}$ &$1.49\pm0.09$ &$0.63\pm0.06$   &$0.47\pm0.05$ \\
$S_{\lambda}/S_K$       & $<3.15\pm0.36$   & $1.32\pm0.19$  & 1.00 \\
\end{tabular}
\caption{Shell parameters, derived from the PSF subtracted
observations, in each waveband.  $P_{\lambda}$ is the per cent
polarization at a radial offset of $1.9$ arcsec from the centre of the
shell. $S_{\lambda}$ is a measure of the shell brightness relative to
the star. The errors on $S_{J}$ and $S_{K}$ assume a 10 per cent error
in the PSF subtraction. The error on $S_{0.55}$ (from which a PSF has
not been subtracted) is based on photon noise.}
\end{table}


\subsubsection{Extinction}
An estimate of the extinction to the star can be made based on the
expected colour for a star of spectral type G5 Ia and the observed
fluxes (HKV89), however there is disagreement over the exact stellar
temperature. Hawkins et al. (1995) found $E(B-V)=0.6$, based on a
photospheric temperature of $T_{\rm eff}=5100$ K which, assuming a
standard interstellar reddening law, leads to $A_{\rm v}=1.8$ mag.
Reddy \& Hrivnak (1999) derive a higher temperature of $T_{\rm
eff}=6750$ K based on an analysis of the optical spectrum. The implied
bluer intrinsic colour leads to a higher estimate of $A_{\rm v}=4.0$
mag. However, in a further spectral analysis, Th\'{e}venin et
al. (2000) dispute this higher temperature and derive $T_{\rm
eff}=5660\pm70$ K. This lower temperature gives a colour excess of
$E(B-V)=0.9$ and, again assuming a standard interstellar
reddening law, an extinction of $A_{\rm v}=2.9$ mag. At a distance of
6 kpc, a significant fraction of this extinction must be interstellar,
and HKV89 quote 2.0 mag for the interstellar component ``assuming a
distance of a few kpc''. We adopt the constraint that the intrinsic
extinction, caused by the circumstellar shell, should be less than 2
magnitudes in the V band.

\subsubsection{Chemistry}
HD 179821 is an O-rich star with ${\rm O/C}=2.6$ (Reddy \& Hrivnak
1999).  The envelope chemistry is also O-rich with OH masers detected
with the same velocity range as the CO emission (Likkel et al. 1991,
1989) and lying in a shell of similar extent to the dust shell
(Claussen 1993). A weak feature at $10$~$\mu$m in the IRAS LRS spectrum
(HKV89) indicates the presence of silicate grains. More recent {\em ISO}
results suggest that the dust is composed mainly of a mixture of
amorphous and crystalline silicates (Waters et al. 1996).
 
\section{Scattering Models}

\subsection{Modelling Procedure}


Details of the Monte Carlo modelling procedure are given in
Appendix~\ref{modap}.  To compare the model results to the
observations, the model images are convolved with a Gaussian profile
of FWHM $0.9$ arcsec, simulating the seeing at the time of the
observations. They are then processed in the same way as the
observations, by forming azimuthally averaged radial profiles. This
also takes advantage of the symmetry in the model to increase the
signal to noise of the computations for a given number of model
photons. The model profiles are normalized to the observations using
the observed polarized flux peak at 1.9 arcsec, since this is not
subject to PSF subtraction errors. We calculate the ratio of the shell
brightness to that of the star for each model and normalize
this to the observations ($S_{\lambda}$ in Table 1) by varying the
dust grain number density, $N_{\rm in}$. This provides a determination of
$N_{\rm in}$ independent of flux calibration, foreground extinction or
knowledge of the spectrum of the star. By assuming a distance, $D$, to
the object, the physical parameters of the model such as optical
depth, $\tau$, and dust mass, $M_{d}$, shown in Table 2 are obtained.

\subsection{Spherically symmetric models}

To simulate a spherically symmetric shell we use a dust density of the
form 
\begin{equation}
N(r) = N_{\rm in}(r/r_{\rm in})^{-\alpha}
\end{equation}
with $\alpha=2$, corresponding to a constant mass-loss rate throughout
the history of the shell. In view
of the O-rich chemistry we initially assume a population of bare
silicate grains.  We treat as free parameters the inner radius of the
shell, $r_{\rm in}$, the dust grain number density at this radius, $N_{\rm
in}$ and the grain size distribution parameter $\beta$ (see Appendix
A). In order to generate the observed degrees of polarization of 30
to 40 per cent after scattering in a realistic geometry such as a
shell, the grains themselves must be able to produce 70 to 80 per cent
polarization. This requires small grains so we initially set
$\beta=6.0$ (AS1, Table 2).

A scattering model using the spherical shell geometry and the AS1
grain model is shown in Fig. 4.  The ring structure, seen in polarized
flux in the observations (Fig. 1), is clearly seen in the model
polarized flux image, but not in the total flux image. This is
because, for lines of sight (LOS) that pass through the inner hollow
region of the shell, scattering occurs predominantly from dust in
front of the star and, therefore, through forward-throwing angles. For
these angles, there is a large contribution to total flux but little
to the polarized flux (which is zero for a LOS through the star). The
effect is to `fill in' the central region of the shell in total flux
but to leave a `hole' in polarized flux. For a spherically symmetric
detached shell, the detached nature of the shell, with a clear
boundary at the inner radius $r_{\rm in}$, will only be evident
observationally (in scattered light) in a polarized flux image.
\begin{figure} 
\label{s50a}
\epsfxsize=14cm \epsfbox[100 -71 771 750]{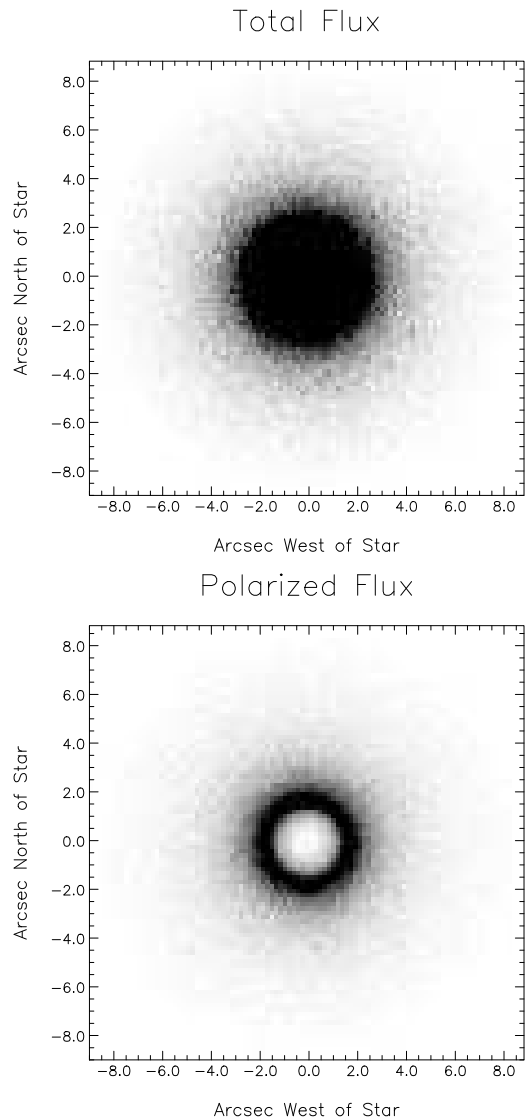}
\vspace{-2cm}
\caption{A spherical shell scattering model (S1) showing the
distribution of total and polarized flux at $1.2$~$\mu$m. The images have
not been smoothed to match the observational data and so retain a
spatial resolution of $0.18$ arcsec. The direct light from the star
occupies a single pixel at the centre of the image and has a total
flux count of 764,000 (in arbitrary units). The images are scaled
between 0 (white) and 50 (black) in the same (arbitrary) units.}
\end{figure}
 
\begin{figure}
\epsfxsize=12cm \epsfbox[110 49 663 780]{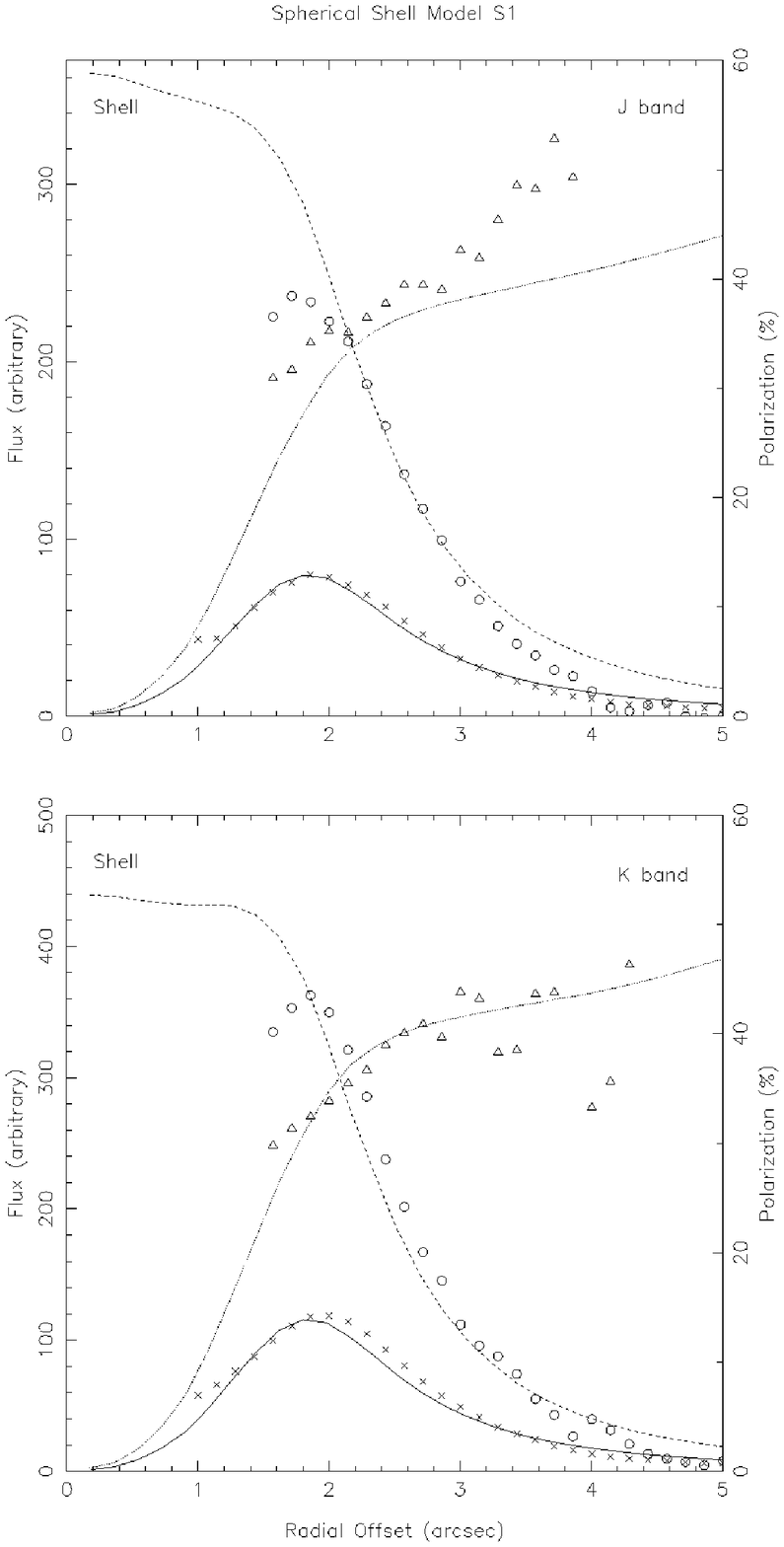}
\caption{A comparison of azimuthally averaged radial profiles through
the spherical shell model S1 with those through the data (Figs. 2 and
3). The error bars, and data points in the central region, have been
omitted for clarity. The model profiles are scaled to the observations
by normalizing the model polarized flux (solid line) to the observed
polarized flux (x) at a radius of $1.9$ arcsec. This also determines
the scaling on the total flux curves (dashed lines). Per cent
polarization is shown as a dotted line.}
\end{figure}
The model profiles are shown in comparison with the observed profiles
in Fig. 5. The observed profiles are plotted using the same symbols as
in Figs. 2 and 3 but with the error bars omitted for clarity. Model
calculations at $1.2$~$\mu$m and $2.2$~$\mu$m are compared with $J$ and $K$ 
band
observations, respectively. The characteristics of the model are
summarized in Table 2.  We find that the S1 model simulates the
observations quite well, suggesting that a spherical shell is a good
approximation to the actual geometry of the CSE. 
The shape of the polarized flux
profile and the position of its peak depend on the scattering
asymmetry parameter, $g$ (see Appendix B), and the radius of the inner
edge of the shell, $r_{\rm in}$. For the S1 model, $g=0.34$ at
$1.2$~$\mu$m (Table B2). If $g$ becomes too large (the grains are too
forward-throwing in their scattering) then the polarized flux peak
moves closer to the star, as the dust in front of the star begins to
dominate the scattering. We find that, in order
to fit the observed polarized flux profile, the dust around HD 179821 must 
have $g<0.6$. 

To produce the high degrees of observed polarization we have
had to use a size distribution with a steep power law ($\beta=6$) so
that a large proportion of small (highly polarizing) grains are
included in the distribution. 
This
produces far too much wavelength dependence in the model and consequently
the shell is much too faint relative to the star at the longer
wavelength ($S_{\rm 1.2}/S_{\rm 2.2}=5.8$ in Table 2). This suggests that we 
must construct a model
which can produce high polarization from larger (greyer)
particles. One possibility is to move to an axisymmetric dust
distribution.

\begin{table*}
\begin{minipage}{150mm}
\caption{A summary of parameters for each model.  Numbers in
parentheses are powers of 10. The AS1 dust models are `astronomical
silicate' with $a_{min}=0.05$~$\mu$m, $a_{max}=2.0$~$\mu$m and $\beta=6.0$.
The AS2 dust model has $\beta=5.5$. $\alpha$, $\epsilon$, $\gamma$ are
parameters of the dust distribution (Section 3.3), $r_{\rm in}$ is the
inner radius of the dust shell in m and $N_{\rm in}$ the grain number
density at that point in m$^{-3}$.  $S_{1.2}/S_{2.2}$ is the ratio of
the relative shell brightness at $\lambda=1.2$~$\mu$m and
$\lambda=2.2$~$\mu$m and a measure of the colour of the
shell. $\tau_{1.2}$ and $\tau_{2.2}$ are the optical depths through
the shell to the star. For the equatorially enhanced models values
along both pole and equator (pole/equator) are given. $M_{d}$ is the
mass of dust in the shell in M$_{\odot}$. The distance, $D$, is
assumed to be 6 kpc.}
\begin{tabular}{|l|l|c|c|c|c|c|c|c|c|c|}
Model  & Dust   & $\alpha$  & $\epsilon$ & $\gamma$ & $r_{\rm in}$ & 
$N_{in}$  & $S_{1.2}/S_{2.2}$ &  $\tau_{1.2}$  & $\tau_{2.2}$  & $M_{d}$   \\
S1     & AS1    &  -2.0    &   0.0      &   0.0    &   1.44(15)   & 
2.70(-1)   &  5.58         & 2.57(-1)         & 7.70(-3)        & 8.20(-2)  \\
D1     & AS2    &  -2.0    &   0.0      &   0.0    &   1.44(15)   &
1.40(-1)   &  4.26         & 2.27(-1)         & 6.68(-2)        & 5.10(-2) \\
D2     & AS2    &  -2.0    &   1.0      &   10.0   &   1.44(15)   &
1.14(-1)   &  4.22         &1.85(-1)/3.69(-1) & 5.44(-2)/1.09(-1) & 5.70(-2) \\
D3     & AS2    &  -2.0    &   2.0      &   3.0    &   1.44(15)   &
7.50(-2)   &  4.10         & 1.22(-1)/3.64(-1)& 3.58(-2)/1.07(-1) & 5.90(-2) \\
D4     & AS2    &  -2.0    &   9.0      &   3.0    &   1.44(15)   &
2.90(-2)   &  3.93         & 4.70(-2)/4.70(-1)& 1.38(-2)/1.38(-1) & 6.60(-2) \\
D5     & AS2    &  -2.0    &   9.0      &   10.0   &   1.44(15)   &
4.60(-2)   &  4.09         & 7.45(-2)/7.44(-1)& 2.19(-2)/2.19(-1) & 7.20(-2) \\
\end{tabular}
\end{minipage}
\end{table*}

\subsection{Equatorially enhanced models}
\begin{figure}
\epsfxsize=13cm \epsfbox[110 59 663 770]{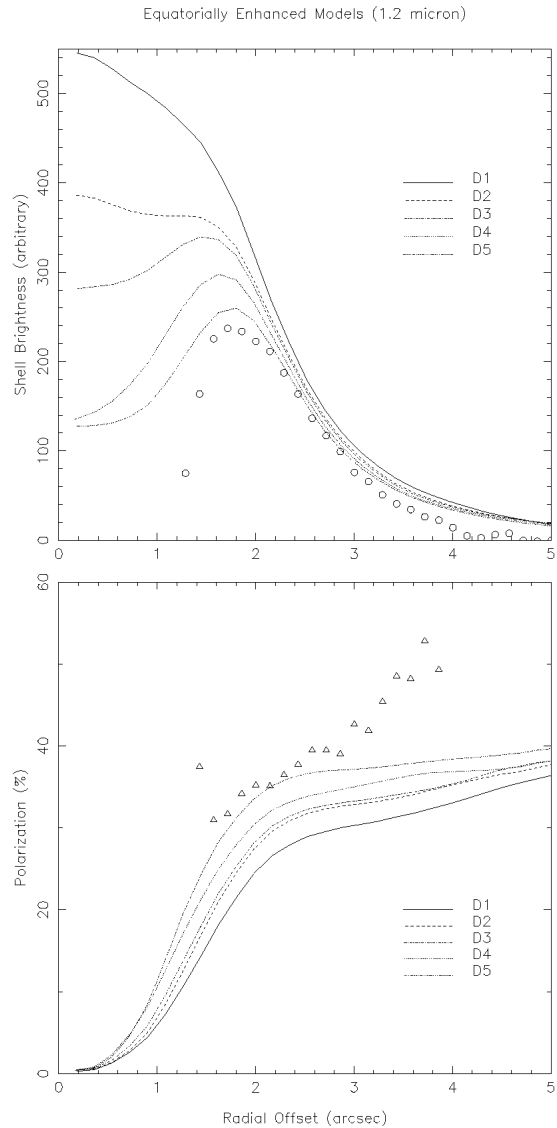}
\caption{Azimuthally averaged radial profiles through
shell models with equatorially enhanced dust distributions, compared with 
the data (Figs. 3). The error bars on the data points
have been omitted for clarity. Profiles of both total flux (upper panel)
and per cent polarization (lower panel) are shown. The parameters for
the models are listed in Table 2.}
\end{figure}
The spherically symmetric geometry results in long lines of sight
through the shell which sample a wide range of scattering
angles. Since maximum polarization is only produced for scattering
angles around $90\degr$ the geometry acts to reduce the maximum
polarization of the AS1 dust model from $\sim80$ per cent to $\sim40$
per cent. By concentrating the dust in the plane of the sky,
scattering angles around $90\degr$ are given more weight and the
polarization can be increased. This can be achieved using a
model with an equatorial density enhancement which is then viewed
pole-on. To simulate this we use a dust
density of the form
\begin{equation}
N(r) = N_{\rm in}(r/r_{\rm in})^{-\alpha}(1+\epsilon\sin^{\gamma}{\theta})
\end{equation}
which is similar to that used to generate axisymmetry in models of
planetary nebulae (e.g. Kahn \& West 1985; Kwok 2000).  The radial
density fall off is now modified to create a density enhancement of
$1+\epsilon$ between the equator and pole. The parameter $\gamma$
controls the way density increases with polar angle, $\theta$, with
higher values of $\gamma$ concentrating more material in the
equatorial region.

In Fig. 6 we show plots of the radial dependence of surface brightness
and polarization at $1.2$~$\mu$m for five `disc' models with various values
of $\epsilon$ and $\gamma$. These models, D1 to D5, simulate
increasing concentrations of dust in the equatorial plane, with D1
being a spherically symmetric model and D5 being the most flattened
with a density contrast (equator to pole) of 10:1 and an angular
dependence of $\sin^{10}\theta$. All are viewed pole-on and have $\alpha=2$ 
with the inner radius, $r_{\rm in}$,
fixed at $1.44\times10^{15}$ m ($D=6$ kpc). A silicate dust model,
AS2, is used which differs from the AS1 model used in Section 3.2 in
that it has a slightly shallower power law for the size distribution
with $\beta=5.5$, thus including larger grains.

Only the D5 model with a 10:1 density contrast
($\epsilon=10$, $\gamma=10$) approaches the observed polarization.
However, increasing the density contrast this much also acts to
flatten the polarization profile so that it no longer matches the
observed rise in polarization with radial offset. 
Increasing the density contrast between equator and poles also causes
the intensity profile to turn over at small radial offsets until, for
the D5 model, there is a peak in intensity similar to that observed in
polarized flux. In other words, there is now a central `hole' in
intensity as well as polarised flux (Fig. 1). There is no evidence for
such a hole or inner shell boundary in the V-band {\em HST} images
(Ueta et al. 2000), although running the D5 simulation at $0.55$~$\mu$m
also produces a central hole. This indicates that any density
enhancement must be slight (D1 and D2 models).

It is also apparent from Table 2 that concentrating the dust in the
equatorial plane does nothing to solve the colour problem -- the shell
is still too blue. The $S_{1.2}/S_{2.2}$ index does decrease slightly
for the more axisymmetric models but this is due to an increase in
extinction for paths in the equatorial plane.  Changes to the
geometry alone cannot increase the polarization produced by the dust
models based on astronomical silicate so as to simultaneously match
the observed polarization and still retain colour neutral scattering at
$J$ and $K$.

\section{Discussion}

The results of the previous section indicate that the radial
distribution of polarized flux, surface brightness and per cent
polarization in the near-IR can be modelled using a spherical distribution
of dust with a sharp inner boundary and an $r^{-2}$ density fall off. It
has not been possible to simultaneously fit the degree of polarization
and the low colour excess ($E(J-K)=-0.3$) since all of the models
considered produce scattering which is far too blue. The shell could
easily be reddened by increasing the optical depth, however, this
would also reduce the polarization. In addition, examination of the V
band {\em HST} image of Ueta et al. (2000) suggests that the shell is
optically thin in the V band, since the surface brightness profile
(Fig. 3) is very similar to that of our $J$ band data. 

So far our dust model has been based on `astronomical silicate'
(Draine 1995) which has been widely used to model reflection nebulae
in star formation regions.
However,
silicates only produce high degrees of polarization for small grains
(close to the Rayleigh limit) which inevitably exhibit strong colour
effects in the scattered intensities, producing predominantly `blue'
reflection nebulae. In Appendix B we tabulate the scattering
properties of various size distributions of `astronomical silicate'
and other materials. For an optically thin
nebula (where absorption is not important), the ratio of scattering cross
sections at $1.2$ and $2.2$~$\mu$m gives a reliable
indication of the nebula colour. In order to produce a nebula with a
ratio of scattering cross sections $C_{\rm sca}(1.2\mu{\rm m})/C_{\rm
sca}(2.2\mu{\rm m})<2.0$, a power law size distribution of astronomical
silicates with $a_{\rm min}=0.05$~$\mu$m and $a_{\rm max}=2.0$~$\mu$m must
have an index $\beta<4.0$. Such a distribution produces very little
polarization and cannot fit our observations. The same is true of the
other forms of silicate listed (olivine, pyroxene) which have similar
refractive indices in the near-IR.  As mentioned in Section 2.2, the
PSF subtraction is only approximate, so that it is possible we have
over-estimated the intrinsic shell polarization, in which case some of
the more highly polarizing grain models (Table B2) may account for the
observations. However, even if we take the raw (before PSF
subtraction) polarization of 20 per cent at $J$ and $K$ as the intrinsic
polarization of the shell, a bare silicate dust model cannot produce
the required colours.
We therefore rule out scattering from spherical silicate particles as
the dominant polarizing mechanism.
  
\subsection{Dust composition}

Some of the emission features seen towards
HD 179821 may be due to metal oxides in the shell, such as
magnesium-iron-oxides (Ueta et al. 2001). There is evidence that these
materials may contribute to the observed structure in the 15--24$~$$\mu$m
spectral region (Henning et al. 1995, Waters et al. 1996). We have
calculated the scattering properties for Fe$_{0.4}$Mg$_{0.6}$O
(Henning et al. 1995) and find that this material produces more
polarization than silicates, for an equivalent size distribution. A
scattering cross section ratio of $R_{1.2/2.2}<2.0$ can be produced
for a power law size index of $\beta<4.5$ (Table B2) which results in
maximum polarization of $<60$ per cent at $1.2$~$\mu$m and $<50$ per cent
at $2.2$~$\mu$m. This is still not sufficient to fit the observed shell
polarization, although it could account for the $\sim20$ per cent
polarization observed in the raw (non-PSF corrected) data. The
material does produce around 10 per cent more polarization at
$1.2$~$\mu$m than at $2.2$~$\mu$m, though, which is contrary to our $J$ and $K$
band observations. In addition, even if metal oxides are present in
the shell, it is not clear that they would be majority contributors to
the scattered light or that they would be distributed so evenly.

The {\em ISO} spectra of HD 179821 also show evidence for crystalline
ice in the form of a $43$~$\mu$m `bump', along with various features
attributable to forms of crystalline silicates (Waters et
al. 1996). 
We have calculated the scattering properties for ice coated
silicates with a coating thickness equal to the grain radius (Table
B2). Ices, especially water ice with a low refractive index, are
highly polarizing so that a coating of ice can increase the
polarization of a silicate grain core significantly (by an amount
increasing with the coating thickness).  However, the scattering
properties of the ice are more wavelength sensitive than silicates, so
that there is a trade off between increasing the maximum polarization
of the dust by applying an ice coating and maintaining a grey
scattering efficiency. Again we have not been able to produce the
required high polarization and grey scattering using ice coated
silicates.


\subsection{Particle size distribution}

The power law size distribution used in the previous section has upper
and lower cut-offs on the grain size so as to keep the number of
grains finite. We have adopted the values $a_{\rm min}$=$0.05$~$\mu$m and
$a_{\rm max}$=$2.0$~$\mu$m, rather than treating them as free model
parameters, and so we should examine these assumed values. For any
size distribution, the particle sizes that contribute most to
scattering will be those that maximize the weighted scattering cross
section, $N(a)C_{\rm sca}(a)$. In order to produce the high
polarizations that we see in the $J$ and $K$ band observations, these
particles must be small, typically with size parameter $x<1$ (or
radius $a<0.2$~$\mu$m at $\lambda=1.2$~$\mu$m), where $x=2\pi a/\lambda$.
This is illustrated in Fig. 7 which shows the polarization produced at
$\lambda=1.2$~$\mu$m as a function of $a$. Particles larger than $x=1$
produce less polarisation and of varying sign (especially for the
dielectric materials such as silicate) which sum to small values when
a distribution of sizes is taken. Since the scattering cross-section
increases with particle radius, $a$, as up to the sixth power, a steep
power law index, $\beta$, is required for the size distribution to
ensure that the smaller (polarizing) particles are present in
sufficient numbers to dominate the scattered light. This in turn means
that the scattering properties are rather insensitive to the upper
size cut-off, $a_{\rm max}$, since there are too few particles of this
size to contribute. At the small size end of the distribution ($x<1$),
scattering is approaching the Rayleigh limit where the cross-section
falls off as $C_{\rm sca}\propto a^{6}$ and polarization is a
maximum. As long as $a_{\rm min}$ is small enough to include Rayleigh
particles, then reducing it further has very little effect on the
polarization or colour of the scattered light. For values of
$\beta<6$, the smallest particles do not contribute to the scattered
flux.
\begin{figure*}
\epsfxsize=18cm \epsfbox[60 19 613 400]{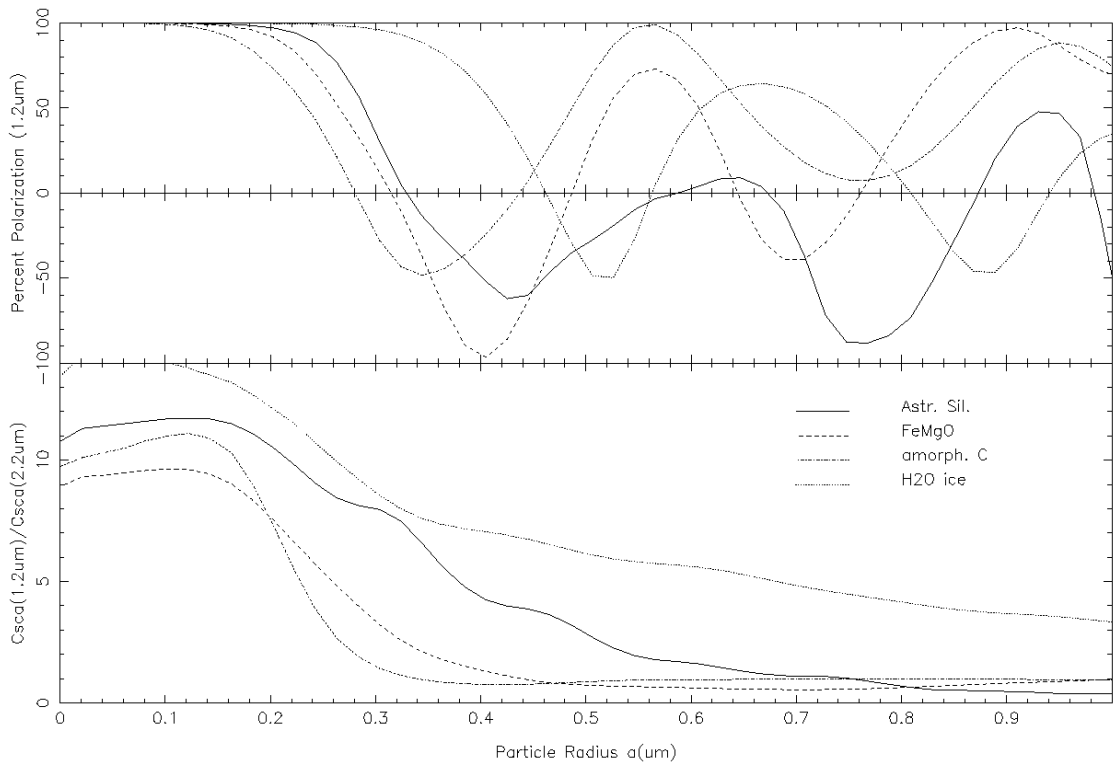}
\caption{Linear polarization at $\lambda=1.2$~$\mu$m (top panel) and the
ratio of scattering cross sections at $\lambda=1.2$~$\mu$m and
$\lambda=2.2$~$\mu$m (bottom panel) as a function of grain radius, $a$,
for various materials. }
\end{figure*}
Fig. 7 shows that it is not possible to generate the observed high
polarization and neutral nebula colour by adjustments to the power
law size distribution, at least for the materials we have
considered. The range of particle sizes responsible for producing high
polarization and neutral colour (low $C_{\rm sca}(1.2\mu{\rm
m})/C_{\rm sca}(2.2\mu{\rm m})$) are mutually exclusive. This would
also be the case for other size distribution functions, such as a
modified power law (Jura 1994) or exponential distribution (Jura
1975).

\subsection{Non-spherical or composite particles}

It is well known that much of the dust in the Galaxy is composed of
non-spherical grains which are responsible for phenomena such as the
interstellar linear and circular polarization (Whittet 1995; Martin
1972). Mid-IR polarization observed in star-forming regions
(e.g. OMC1, Aitken et al. 1997) shows that aligned, non-spherical
particles are present in circumstellar environments also. Given the
wide-spread distribution of non-spherical grains it seems reasonable
to assume that at least a proportion of the dust at its point of
origin, in the circumstellar envelopes of evolved stars, will
also be non-spherical.

The scattering properties of highly spheroidal particles differ
significantly from spheres and can show a combination of the
characteristics of larger and smaller particles.  Zakharova \&
Mishchenko (2000) find that for spheroids with aspect ratios of 20:1,
the polarization reaches 100 per cent at a scattering angle of 90
\degr, typical of Rayleigh scattering, even for particles with size
parameters $x=3.5$. However, the scattering efficiency as a function
of size parameter is more typical of that of larger spheres, showing a
weak dependence on $x$. It therefore seems plausible that the high
polarizations and weak dependence of $C_{\rm sca}$ on grain size,
required to explain our observations, may be satisfied if the
scatterers are spheroidal with large aspect ratios (needle-like or
plate-like). An alternative to highly elongated particles could be
provided by composite or aggregate particles where the overall size of
the particle is comparable to the wavelength, but many of the
scattering properties (such as the polarization) are characteristic of
the component sizes. Such behaviour has been seen in low density
aggregates of spheres (West 1991).

\subsection{Dust mass, extinction and the distance to HD 179821}

The dust optical depths shown in Table 2 have been obtained by fitting
to the scattered surface brightness of the shell relative to the
star, and are intrinsic to the object and not dependent on the assumed
distance. If the object is closer than 6 kpc then path lengths through
the shell would decrease and the dust density would have to be
increased to maintain the observed shell brightness (and hence
optical depth). So our model fits do not directly constrain the distance
estimates.
Using the silicate dust model and spherical (S1)  geometry we
calculate an extinction to the star at $0.55$~$\mu$m (roughly V band) of
$\tau_{\rm 0.55}=1.67$. This is consistent with estimates of the 
extinction to the star in the V band of between  1 and 2 mag 
(Section 2.3.3). 

For the S1 model we calculate a dust mass of $M_{d}=0.08$~M$_{\odot}$,
assuming an outer shell radius of 9 arcsec (Kastner \& Weintraub
1995), a distance of 6 kpc and a density of $3.0\times10^{3}$
~kg~m$^{-3}$ for silicate dust. This is of the same order as estimates
of the dust mass obtained from mid-IR modelling of $0.03$~M$_{\odot}$
(Jura \& Werner 1999) and $0.04$~M$_{\odot}$ (Hawkins et
al. 1995). These authors assume values for the gas-to-dust mass ratio
of between 100 and 160 which gives a total (gas+dust) mass in our
model of between 8 and 13~M$_\odot$.  Our estimate of $M_{d}$ is
sensitive to the dust model, in particular the scattering
cross-section, and could easily be smaller, in agreement with the
estimates from mid-IR emission. For example, increasing the scattering
cross-section by using slightly larger grains (D1 in Table 2) gives
$M_{d}=0.05$~M$_{\odot}$, although this no longer fits the observed
polarizations.

\section{Conclusions}

The circumstellar envelope of the evolved star HD 179821 is seen by
scattered light in the near-IR and appears as a detached shell in
linearly polarized light. We have used observations of a PSF
calibration star to subtract the atmospherically scattered PSF of HD
179821 in order to obtain an image of the shell and hence calculate
its intrinsic linear polarization. We find high degrees of
polarization, between 30 and 40 per cent, in both $J$ and $K$ bands,
with evidence for a gradual increase in polarization with radial
offset from the star.
The colours are remarkably neutral for scattering, with a
colour excess in the shell of $E(J-K)=-0.3$.

We use an axisymmetric scattering code to model the $J$ and $K$ band
observations and find that the distribution of polarized and total
flux can be produced by a spherically symmetric shell of
dust with an $r^{-2}$ density distribution. The model shows 
that, when scattering from
such a spherically symmetric shell, the detached nature of the shell is
only apparent in polarized flux, with the total flux image showing no
evidence for an inner boundary. By fitting the models to the observed
brightness of the shell relative to the star, we
obtain an optical depth to the star at $1.2$~$\mu$m of $\tau_{1.2}=0.26$
which is consistent with estimates of the V band extinction. Assuming a
distance to HD 179821 of 6 kpc, we derive an inner shell radius of
$r_{in}=1.44\times10^{15}$~m, a dust number density of
$N_{in}=2.70\times10^{-1}$~m$^{-3}$ at $r_{in}$ and a total mass of
dust in the shell of $M_{d}=0.08$~M$_{\odot}$, assuming a power-law size
distribution of silicate grains.  These estimates are sensitive to
uncertainties in the dust model but the dust mass is of the same order
as that derived from mid-IR imaging of the warm dust.

We have experimented with equatorially enhanced density distributions,
viewed pole-on, but conclude that if the shell does have an increased
concentration of dust in the equatorial plane then the contrast
between equator and poles cannot be much more than 2:1, otherwise the
radial dependence of linear polarization becomes too flat to fit the
observations. 

Although the spherical shell model fits the observed distributions of
total and polarized flux quite well, we have not been able to
simultaneously account for the high degrees of linear polarization and
the unusually `grey' scattering at $J$ and $K$. We attribute this to
inadequacies in the dust model and have explored various
materials such as silicates, carbons, metal oxide and also ice-coated
particles. If the high intrinsic polarizations implied by our PSF
subtraction are correct, then none of the materials has reproduced the
observed colours and polarizations and a distribution of homogeneous
spherical particles, as assumed here, is unlikely to do so. Instead,
we suggest that the dust surrounding HD 179821 may consist of either
highly elongated needle- or plate-like particles or of composite
grains with many smaller inclusions.

It would be very useful to obtain higher spatial resolution imaging
polarimetry observations (such as coronographic 8-m data) and/or
spectropolarimetry over the optical and near-IR, to ascertain the
polarization in the shell more accurately and to determine the
scattering behaviour of the dust over a wider wavelength range.

\section*{Acknowledgments}
We are grateful to Fran\c{c}ois M\'{e}nard for allowing us to use and
modify his original Monte Carlo scattering code. Phil Lucas is thanked
for assistance with scattering from coated spheres. We
thank the NSCA Astronomy Digital Image Library (ADIL). Use was made of the 
{\tt SIMBAD}
astronomical database. The modelling in this paper was based on data
taken at the United Kingdom Infrared Telescope. An anonymous referee
is thanked for helpful comments.

\section*{References}
\begin{tabbing}
~~~~~~~~~~\= \\
Aitken D.K., Smith C.H., Moore T.J.T., Roche P.F., \\
  \> Fujiyoshi T., Wright C.M., 1997, MNRAS, 286, 85 \\
Bohren C.F., Huffman D.R., 1983, Absorption and Scattering \\
  \> of Light by Small Particles, John Wiley \& Sons, \\ 
  \> New York \\
Claussen M.J., 1993 in Astrophysical Masers, ed. A.W. Clegg, \\
  \> G.E. Nedoluha, Spiringer, Berlin, p. 353 \\
Dorschner J., Begemann B., Henning Th., J$\ddot{\rm a}$ger C., \\
  \> Mutschke H., 1995, A\&A, 300, 503 \\
Draine B.T., 1985, ApJS, 57, 587 \\
Gledhill T.M., Chrysostomou A., Hough J.H., Yates J.A., \\
  \> 2001, MNRAS, 322, 321 \\
Hawkins G.W., Skinner C.J., Meixner M.M., Jernigan J.G., \\
  \> Arens J.F., Keto E., Graham J.R., 1995, ApJ, 452, 314 \\
Henning Th., Begemann B., Mutschke H., Dorschner J., \\
  \> 1995, A\&AS, 112, 143 \\
Hrivnak B.J., Kwok S., Volk K.M., 1989, ApJ, 346, 265 \\
Jura M., 1994, ApJ, 434, 713 \\
Jura M., Werner M.W., 1999, ApJL, 525, L113 \\
Kahn F.D., West K.A., 1985, MNRAS, 212, 837 \\
Kastner J.H., Weintraub D.A., 1995, ApJ, 452, 833 \\
Kwok S., 2000, in The Origin and Evolution of Planetary \\
   \> Nebulae, Cambrige Astrophysics Series 31, Cambridge \\
   \> University Press \\
Likkel L., 1989, ApJ, 344, 350 \\
Likkel L., Forveille T., Omont A., Morris M., 1991, \\
  \> A\&A, 246, 153 \\
Martin P.G., 1972, MNRAS, 159, 179 \\
M$\acute{\rm e}$nard F., 1989, Ph.D. Thesis, University of Montreal \\
Preibisch Th., Ossenkopf V., Yorke H.W., Henning Th., \\
  \> 1993, A\&A, 279, 577 \\
Reddy B.E., Hrivnak B.J., 1999, AJ, 117, 1834 \\
Th$\acute{\rm e}$venin F., Parthasarathy M., Jasniewicz G., 2000, \\
   \> A\&A, 359, 138 \\
Ueta T., Meixner M., Bobrowsky M., 2000, ApJ, 528, 861 \\
Ueta T., Speck A.K., Meixner M., Dayal A., Hora J.L., \\
   \> Fazio G., Deutsch L.K., Hoffmann W.F., 2001, in eds. \\
   \> Szczerba R., Tylenda R., Gorny S.K., Post-AGB Objects \\ 
   \> as a Phase of Stellar Evolution, in press \\
Warren S.G., 1984, Applied Optics, 23, 1206 \\
Waters L.B.F.M., Molster F.J., de Jong T., Beintema D.A., \\
   \> et al., 1996, A\&A, 315, L361 \\ 
West R.A., 1991, Applied Optics, 30, 5316 \\
Whittet D.C.B., 1995, in Greenberg J.M., ed., The Cosmic \\
   \> Dust Connection, Kluwer, Dordrecht, p. 155 \\
Zakharova N.T., Mishchenko, M.I., 2000, Applied Optics, \\
   \> 39, 5052 \\
Zuckerman B., Dyck H.M., 1986, ApJ, 311, 345 \\
\end{tabbing}

\begin{appendix}

\section{Details of the Monte Carlo scattering model}
\label{modap}
The scattering models are computed using a modified version of the
Monte Carlo scattering code of M\'{e}nard (1989).  A given number of
photons (usually $10^{7}$ for the optically thin models) are followed
as they leave the star and propagate through the dusty envelope. At
each scattering event, the new photon path and Stokes intensities are
calculated.  The density structure of the envelope can be spherically
symmetric or axisymmetric (2D) and is specified as a table of grain
number density at each point ($r$, $\theta$) where $r$ is the radial
distance from the star and $\theta$ is the polar angle ($\theta=0$ for
a pole-on view and $\theta=90$ for an equatorial view). Photons
emerging from the nebula are collected in bins according to their
polar angle to form a 2-D scattered light image of the nebula for each
viewing inclination angle. In practice these bins are averaged over
inclination angle, to improve signal to noise, so that the bin
approximating a pole-on view (bin 1) actually encompasses viewing
angles in the range $\theta=0\rightarrow10\degr$. Since the
observations of HD 179821 suggest that the nebula geometry is either
spherically symmetric, or axisymmetric but viewed pole-on, this bin is
used throughout. 

The dust grain model is specified by the refractive index of
the dust, $m$, and the grain size distribution. We assume a truncated
power law distribution of grain sizes such that the number of grains
of size $a$ per unit volume is given by $n(a) = Aa^{-\beta}$ where
$a_{\rm min} \le a \le a_{\rm max}$ and $A$ is a constant defined such
that
\begin{equation}
N = A \int^{amax}_{amin}a^{-\beta}da
\end{equation}
and $N$ is the total number of grains per unit volume, which is given
at each ($r$,$\theta$) position by the density table. The grains are
assumed to be spherical and the scattering calculations use Mie
theory.

The model geometry suggested by the observations is a hollow shell,
with inner radius $r_{\rm in}$ and outer radius $r_{\rm out}$. The
number density of dust grains at the inner radius, $r_{\rm in}$, is
$N_{\rm in}$, and outside the shell ($r < r_{\rm in}$ and $r > r_{\rm
out}$) it is zero. Within the shell the grain number density is given
by the density table. We set the outer shell boundary to 
$5.4\times10^{4}$~AU, corresponding to 9 arcsec at 6 kpc, as indicated 
by the extent of the scattered light detected by Kastner \& Weintraub
(1995). The model images at each viewing angle are
computed on a 2D grid of $99\times99$ pixels, with the star at the
central pixel. For modelling purposes we have ignored the observed
offset of $0.3$ arcsec between the star and the centre of the shell.
With an outer shell radius of $r_{\rm out}=9$ arcsec
then a model pixel is equivalent to $0.18$ arcsec. 

\section{Scattering properties of various dust grain models}

We list here the results of scattering calculations for various
particle size distributions. All calculations, assume Mie scattering
and are based on modified versions of the {\tt BHMIE} and {\tt BHCOAT}
spherical particle algorithms of Bohren \& Huffman (1983). Power-law
size distributions (Section 3.1) are used throughout with $a_{\rm
min}=0.05$~$\mu$m and $a_{\rm max}=2.0$~$\mu$m. Table B1 lists optical
constants at $\lambda=1.2$~$\mu$m and $\lambda=2.2$~$\mu$m for various
materials and Table B2 lists their scattering properties.

The albedo, $\omega$ is the ratio of scattering and extinction
cross-sections. The asymmetry parameter, $g$, is the average cosine
of the scattering angle and a measure of the forward-throwing nature
of the scattering. For a particle which scatters light equally in all
directions, $g=0$, whereas for forward-throwing scattering, $g$ lies
in the range $0 < g \le 1$. More formal definitions may 
be found in Bohren \&
Huffman (1983).

\begin{table*}
\begin{minipage}{150mm}
\caption{Optical constants for various materials at $\lambda=1.2\mu$m and 
$\lambda=2.2\mu$m.}
\label{dust}
\begin{tabular}{|l|c|c|c|c|c|p{2in}|}
Material        & $m (\lambda=1.2\mu$m) & $m (\lambda=2.2\mu$m) & Comment                 & Ref \\
Ast Sil      & 1.71, 0.03    & 1.71, 0.03 & `astronomical silicate' & 1   \\
Mg$_{0.8}$Fe$_{1.2}$SiO$_{4}$      
             & 1.790, 0.103  & 1.866, 0.052 & Fe-rich olivine glass & 2   \\
Mg$_{0.6}$Fe$_{0.4}$SiO$_{3}$
             & 1.640, 0.004  & 1.620, 0.002 & Fe-rich pyroxene glass & 2  \\
Fe$_{0.4}$Mg$_{0.6}$O & 1.95, 0.31 & 2.15, 0.26 & iron-magnesium oxide & 3 \\
Amor. Carbon & 2.51, 0.76   & 2.79, 0.80 & amorphous carbon        & 4   \\
H$_{2}$O Ice & 1.298, 6.7(-6) & 1.262, 2.6(-4)& water ice at 140 K  & 5   \\
\end{tabular}
\\
References: 1 Draine 1985; 2 Dorschner et al. 1995; 3 Henning et al. 1995;
4 Preibisch et al. 1993; 5 Warren 1984.
\end{minipage}
\end{table*}

\begin{table*}
\begin{minipage}{150mm}
\caption{Scattering properties for power-law size distributions of spherical
particles. $\beta$ is the power-law size distribution index, $C_{sca}(1.2)$ 
and $C_{sca}(2.2)$ are the scattering cross-sections at $1.2$ and $2.2$~$\mu$m,
$\omega_{1.2}$ and $\omega_{2.2}$ the albedos, $g_{1.2}$ and $g_{2.2}$ the 
scattering asymmetry  parameters,  $P_{m}(1.2)$ and $P_{m}(2.2)$ the
maximum linear polarizations and R$_{1.2/2.2}$ is the ratio
$C_{sca}(1.2)/C_{sca}(2.2)$.}
\begin{tabular}{|l|c|c|c|c|c|c|c|c|c|c|}
Material &$\beta$ & $C_{sca}(1.2)$ & $C_{sca}(2.2)$ & $\omega_{1.2}$ 
         & $\omega_{2.2}$ 
      & $g_{1.2}$ & $g_{2.2}$ & $P_{m}(1.2)$ & $P_{m}(2.2)$ & R$_{1.2/2.2}$ \\
Ast Sil & 6.0   & 5.27(-16)    & 9.07(-17)    & 0.62         &  0.37        
        & 0.34    & 0.32    & 75         & 77         & 5.81                \\
        & 5.5   & 1.02(-15)    & 2.29(-16)    & 0.71         &  0.54      
        & 0.41    & 0.40    & 63         & 65         & 4.47                \\
        & 5.0   & 2.15(-15)    & 6.35(-16)    & 0.78         &  0.69
        & 0.47    & 0.46    & 50         & 51         & 3.39                \\
        & 4.5   & 4.88(-15)    & 1.90(-15)    & 0.83         &  0.79
        & 0.52    & 0.51    & 35         & 37         & 2.57                \\
        & 3.5   & 3.02(-14)    & 1.96(-14)    & 0.83         & 0.85
        & 0.60    & 0.57    & $<10^{\dagger}$ & $<10^{\dagger}$  & 1.54     \\
Mg$_{0.8}$Fe$_{1.2}$SiO$_{4}$ 
        & 6.0   & 5.69(-16)    & 1.22(-16)    & 0.36         &  0.35       
        & 0.33    & 0.31    & 77         & 71         & 4.68                \\
        & 5.5   & 1.06(-15)    & 2.97(-16)    & 0.45         &  0.50
        & 0.39    & 0.38    & 68         & 57         & 3.56                \\
        & 5.0   & 2.12(-15)    & 7.90(-16)    & 0.54         &  0.64
        & 0.46    & 0.43    & 57         & 45         & 2.69                \\
        & 4.5   & 4.53(-15)    & 2.22(-15)    & 0.59         &  0.72
        & 0.52    & 0.47    & 44         & 30         & 2.04                \\
        & 3.5   & 2.53(-14)    & 2.04(-14)    & 0.64         & 0.78
        & 0.62    & 0.53    & 20         & $<10$      & 1.24                \\
Mg$_{0.6}$Fe$_{0.4}$SiO$_{3}$
        & 6.0   & 4.66(-16)    & 7.61(-17)    &  0.91         &  0.87
        & 0.35    & 0.33    & 76         & 78         & 6.12                \\
        & 5.5   & 9.31(-16)    & 1.98(-16)    &  0.94         &  0.94
        & 0.42    & 0.42    & 67         & 64         & 4.70                \\
        & 5.0   & 2.04(-15)    & 5.75(-16)    &  0.96         &  0.97 
        & 0.48    & 0.48    & 50         & 53         & 3.55                \\
        & 4.5   & 4.83(-15)    & 1.80(-15)    &  0.97         &  0.98
        & 0.53    & 0.53    & 35         & 39         & 2.68                \\
        & 3.5   & 3.31(-14)    & 2.06(-14)    &  0.97         &  0.99
        & 0.59    & 0.59    &  $<10^{\dagger}$ &  $<10^{\dagger}$ & 1.61    \\
Fe$_{0.4}$Mg$_{0.6}$O 
        & 6.0   & 6.71(-16)     & 1.45(-16)    & 0.20         &  0.14
        & 0.29    & 0.27    & 82         & 77         & 4.62                \\
        & 5.5   & 1.16(-15)     & 3.27(-16)    & 0.26         &  0.22
        & 0.35    & 0.33    & 77         & 69         & 3.55                \\
        & 5.0   & 2.17(-15)     & 8.00(-16)    & 0.33         &  0.32
        & 0.42    & 0.39    & 68         & 58         & 2.71                \\
        & 4.5   & 4.34(-15)     & 2.14(-15)    & 0.39         &  0.41
        & 0.48    & 0.44    & 59         & 47         & 2.02                \\
        & 3.5   & 2.17(-14)     & 1.66(-14)    & 0.47         &  0.51
        & 0.62    & 0.54    & 39         & 27         & 1.31                \\
Amo Car & 6.0   & 1.11(-15)    & 2.16(-16)    & 0.20         &  0.11
        & 0.22    & 0.20    & 82         & 81         & 5.14                \\
        & 5.5   & 1.80(-15)    & 4.50(-16)    & 0.25         &  0.17
        & 0.26    & 0.24    & 76         & 73         & 3.99                \\
        & 5.0   & 3.08(-15)    & 1.01(-15)    & 0.30         &  0.24
        & 0.31    & 0.28    & 68         & 66         & 3.06                \\
        & 4.5   & 5.66(-15)    & 2.42(-15)    & 0.35         &  0.32
        & 0.37    & 0.33    & 60         & 48         & 2.34                \\
        & 3.5   & 2.46(-14)    & 1.71(-14)    & 0.44         &  0.44 
        & 0.52    & 0.45    & 44         & 42         & 1.45                \\
H$_{2}$O ice
        & 6.0   & 1.20(-16)    & 1.56(-17)    & 1.0        &  0.90
        & 0.40    & 0.36    & 94         & 97         & 7.68                \\
        & 5.5   & 2.63(-16)    & 4.30(-17)    & 1.0          &  0.95   
        & 0.52    & 0.48    & 90         & 94         & 6.12                \\
        & 5.0   & 6.63(-16)    & 1.37(-16)    & 1.0          & 0.98
        & 0.62    & 0.59    & 83         & 90         & 4.86                \\
        & 4.5   & 1.89(-15)    & 4.80(-16)    & 1.0          & 0.99
        & 0.70    & 0.67    & 74         & 85         & 3.93                \\
        & 3.5   & 1.97(-14)    & 7.11(-15)    & 1.0          & 1.0
        & 0.79    & 0.77    & 46         & 72         & 2.77                \\
Mg$_{0.8}$Fe$_{1.2}$SiO$_{4}$ + H$_{2}$O ice
        & 6.0   & 1.72(-16)    & 2.68(-17)    & 0.55
        & 0.43          & 0.34    & 0.31    & 91    & 90    & 6.42    \\
        & 5.5   & 3.60(-16)    & 7.29(-17)    & 0.67
        & 0.62          & 0.43    & 0.40    & 85    & 84    & 4.94    \\
        & 5.0   & 8.45(-16)    & 2.23(-16)    & 0.76  
        & 0.78          & 0.51    & 0.48    & 77    & 77    & 3.80   \\
        & 4.5   & 2.20(-15)    & 7.41(-16)    & 0.82
        & 0.86          & 0.58    & 0.54    & 67    & 68    & 2.97    \\ 
        & 3.5   & 1.92(-14)    & 9.58(-15)    & 0.87        
        & 0.92          & 0.68    & 0.62    & 48    & 47    & 2.00    \\
\end{tabular}
\\
$\dagger$ a negatively polarised backscattering peak is also produced \\
\end{minipage}
\end{table*}

\end{appendix}

\end{document}